\documentclass[conference]{IEEEtran}
\IEEEoverridecommandlockouts

\usepackage{amsmath}
\usepackage{graphicx}
\usepackage{amssymb}
\usepackage{cases}
\usepackage{cite}
\usepackage[ruled,vlined,lined,ruled,linesnumbered]{algorithm2e}
\usepackage{relsize}
\usepackage[nodisplayskipstretch]{setspace}
\usepackage[usenames, dvipsnames]{color}
\usepackage{epstopdf}
\usepackage[font=small,skip=0pt]{caption}
\begin{document}
\title{User Cooperation for Enhanced Throughput Fairness in Wireless Powered Communication Networks}

\author{Mingquan~Zhong,~Suzhi~Bi, and Xiaohui~Lin\\
College of Information Engineering, Shenzhen University,\\ Shenzhen,  Guangdong, China 518060\\ E-mail:~zhongmingquan@email.szu.edu.cn, \{bsz,~xhlin\}@szu.edu.cn \vspace{-2ex}
      \thanks{This work was supported in part by the National Natural Science Foundation of China (project number 61501303 and 61171071), the Guangdong Natural Science Foundation (project munber 2015A030313552), and the Foundation of Shenzhen City (project number GJHS20120621143440025 and ZDSY20120612094614154).} }

\maketitle
\begin{abstract}
This paper studies a novel user cooperation method in a wireless powered communication network (WPCN), where a pair of distributed terminal users first harvest wireless energy broadcasted by one energy node (EN) and then use the harvested energy to transmit information cooperatively to a destination node (DN). In particular, the two cooperating users exchange their independent information with each other to form a virtual antenna array and transmit jointly to the DN. By allowing each user to allocate part of its harvested energy to transmit the other's information, the proposed cooperation can effectively mitigate the user unfairness problem in WPCNs, where a user may suffer from very low data rate due to the poor energy harvesting performance and high data transmission consumptions. We derive the maximum common throughput achieved by the cooperation scheme through optimizing the time allocation on wireless energy transfer, user message exchange, and joint information transmissions. Through comparing with some representative benchmark schemes, our results demonstrate the effectiveness of the proposed user cooperation in enhancing the throughput performance under different setups.
\end{abstract}

\IEEEpeerreviewmaketitle

\section{Introduction}
Wireless communication devices are conventionally powered by batteries, which have to be replaced/recharged once the energy is depleted. In practice, frequent manual battery replacement/recharging could be inconvenient and often costly especially in large-size wireless network (e.g., wireless sensor network (WSN) for environment monitoring). Besides, it leads to frequent communication outage that degrades the quality of service. Alternatively, radio frequency (RF) enabled wireless energy transfer (WET) technology enables the wireless devices to continuously harvest energy from RF signals to achieve self-sustainable wireless network operation \cite{2014:Bi,2015:Lu,2014:Krikidis,2016:Bi}.

One interesting application of WET is wireless powered communication network (WPCN), where wireless devices transmit information using the power harvested by means of WET\cite{2013:Zhou,2014:Ju1}. For instance, \cite{2014:Ju1} proposed a harvest-then-transmit protocol in WPCN, where one hybrid access point (HAP) with single-antenna first broadcasts RF energy to all the users in the downlink, and then the users transmit their information to the HAP in the uplink using their individually harvested energy in a time-division-multiple-access (TDMA) manner. However, such design may induce serious user unfairness, named doubly-near-far problem, where users far away from the HAP achieve very low throughput because they suffer from both poor energy harvesting performance and high data transmission consumptions. To enhance the user fairness, \cite{2014:Ju3} proposed a two-user cooperation scheme where the near user helps relay the far user's information to the HAP. \cite{2015:HeCHen} extended \cite{2014:Ju3} to a multi-relay scenario and proposed a harvest-then-cooperate protocol to coordinate the transmissions of nearby users to forward the message of a far-away user. Both works consider using a HAP to transmit energy and receive information, which, however, is the essential cause of the doubly-near-far problem. To further enhance user fairness, separately located energy and information access points (APs) are considered to more flexibly balance the energy and information transmissions, as now the poor energy harvesting performance of a WD can be compensated by low information transmit power to a nearby information AP \cite{2014:Huang1,2014:Lee1,{2015:Bi:Placement Optimization}}.

In this paper, we present a novel user cooperation method in WPCN with separately located energy and information APs as shown in Fig.~1. Specifically, the two energy-harvesting users $X$ and $Y$ exchange their individual messages with each other to form a virtual antenna array, and transmit jointly to the DN. The key contributions of this paper are summarized as follows:
\begin{itemize}
  \item We present a new user cooperation method for enhancing throughput fairness in WPCN. Compared to the existing cooperation scheme where one user acts as the relay for the other, the proposed method allows the users to share their harvested energy and transmit jointly, thus achieving both energy diversity and channel diversity gains.
  \item We derive the maximum common throughput achieved by the cooperation scheme through optimizing the time allocation on wireless energy transfer, user message exchange, and joint information transmissions in a fixed transmission time slot.
  \item We perform numerical analysis to study the impact of system setups to the performance of the proposed cooperation method. Through comparisons with other benchmark schemes, we show that the proposed cooperation can effectively improve the throughput performance, especially when the inter-user channels are sufficiently strong to support efficient information exchange and the two users have comparable user-to-DN channels.

\end{itemize}

\begin{figure}
  \centering
   \begin{center}
      \includegraphics[width=0.4\textwidth]{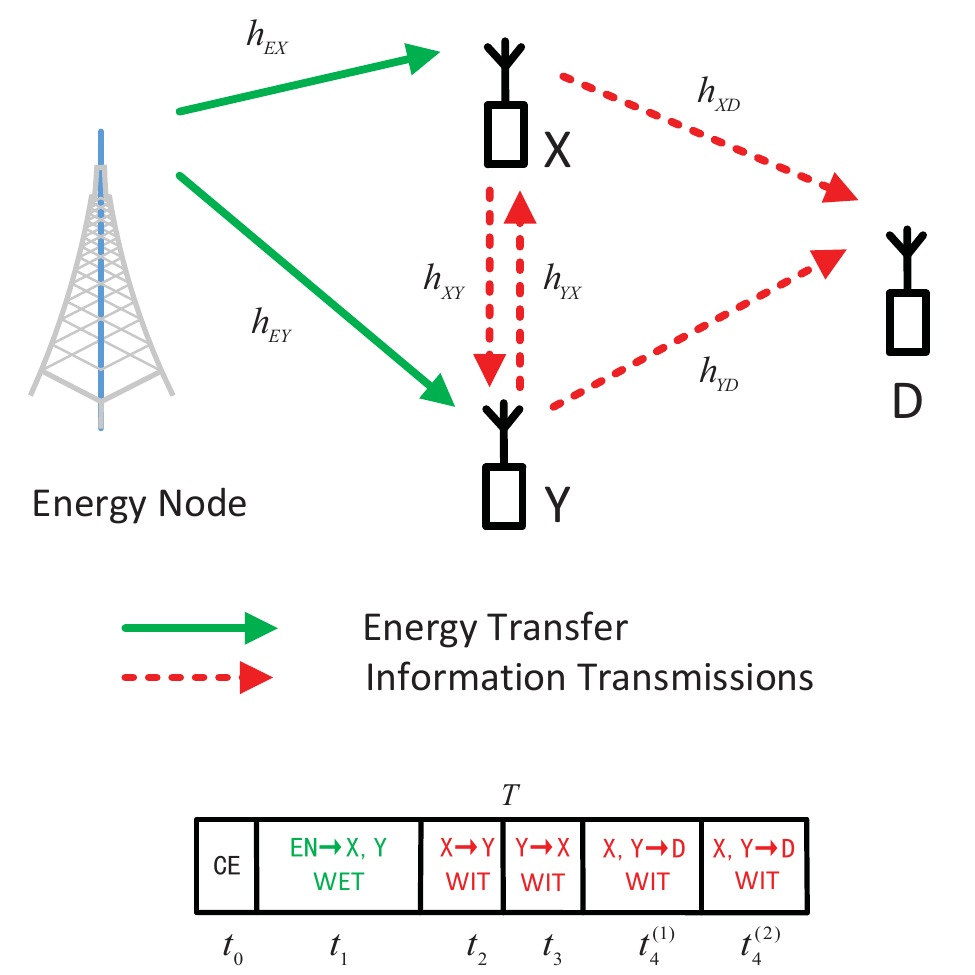}
   \end{center}
  \caption{The proposed user cooperation method and the operating protocol. }
  \label{Fig.1}
\end{figure}

\section{System Model}
As shown in Fig.~1, we consider a WPCN with two users $X$ and $Y$ who first harvest RF energy from the EN and then transmit cooperatively their data to the DN. The EN is assumed to have a constant energy supply and both terminal users have no other embedded energy source, thus need to store the harvested energy in a rechargeable battery for information transmission to the DN. It is assumed that each node is equipped with single antenna. For each user, the antenna is used for both energy harvesting and communication in a time-division-duplexing (TDD) manner \cite{2013:Zhou}. The EN also has a similar TDD circuit structure to switch between energy transfer and communication, e.g, for performing channel estimation, with the WDs.

It is assumed that all the channels are reciprocal and under quasi-static flat-fading, where the channel gains remain constant during each transmission block of duration $T$ but vary from one block to another. At the beginning of a transmission block channel estimation (CE) is performed within a fixed duration $t_0$. Then, in the remainder of a tagged transmission block, $t_1$ amount of time is assigned for WET while the remaining time is assigned for WIT. In the next two time slots with duration $t_2$ and $t_3$, respectively, the two users exchange with each other their own messages. In the last time slot of length $t_4$, the two users transmit jointly their information to the DN. Specially, $t_4^{(1)}$ amount of time is allocated to transmit user $X$'s information, and the rest of the time slot with duration $t_4^{(2)}$ is for transmitting $Y$'s information, with $t_4=t_4^{(1)}+t_4^{(2)}$. Note that we have a total time constraint
\begin{equation}
\small
t_0+t_1+t_2+t_3+t_4^{(1)}+t_4^{(2)}=T.
\end{equation}
For the simplicity of exposition, we set without loss of generality $T=1$.

The notations of channel gains are given in Fig.~1. In the CE stage, user $X$ and $Y$ broadcast their pilot signals, so that EN has the knowledge of $h_{EX}$ and $h_{EY}$, the DN knows $\alpha_{_{XD}}$ and $\alpha_{_{YD}}$, and user $X$ ($Y$) knows $\alpha_{_{YX}}$ ($\alpha_{_{XY}}$), respectively, where $\alpha_{_{XY}}$ denotes the complex channel coefficient between $X$ and $Y$ with $h_{XY}\triangleq |\alpha_{XY}|^2$. Then, each node feeds back their known CSI to a control point, which calculates and broadcasts the optimal time allocation $(t_1^*,t_2^*,t_3^*,t_4^{(1)*},t_4^{(2)*})$ to all the nodes in the network. Notice that under this setup, user $X$ and $Y$ have no knowledge of $\alpha_{_{XD}}$ and $\alpha_{_{YD}}$. Therefore, transmit beamforming is not applicable at user $X$ and $Y$.

In the first time slot, we let $P_t$ denote the transmission power of the EN and assume that the energy harvested from noise is negligible by the users. Then, the amount of energy harvested by user $X$ and $Y$ can be expressed as \cite{2014:Ju1}
\begin{equation}
\small
  E_X={\eta}{t_1}{P_t}{h_{EX}},\  E_Y={\eta}{t_1}{P_t}{h_{EY}},
\end{equation}
where $0<\eta<1$ is the energy harvesting efficiency assumed equal for both users.

In the subsequent WIT stage, we assume that both user $X$ and $Y$ exhaust the harvested energy, and each transmits with a constant power during the WIT stage. Then, the transmit power of $X$ and $Y$ is $P_X=E_X/(t_2+t_4)$ and $P_Y=E_Y/(t_3+t_4)$, respectively. Let $S_X(t)$ denote the baseband signal of the user $X$ transmitted in $t_2$ with $ E[|S_X(t)|^2] = 1$, the received signal at user $Y$ is then expressed as
\begin{equation}
\small
Z_Y^{(2)}(t)=\sqrt{P_X}{\alpha}_{_{XY}}S_X^{(2)}(t)+n_Y^{(2)}(t),
\end{equation}
where $t\in(t_0+t_1,t_0+t_1+t_2]$, and $n_Y^{(2)}(t)$ denotes the receiver noise at user $Y$. Without loss of generality, we assume the receiver noise power is $N_0$ at all receivers. Then, user $Y$ can decode the $X$'s information at a rate given by
\begin{equation}
\label{Rx2}
\small
R_X^{(2)}=t_2\log_{2}\left(1+\frac{E_Xh_{XY}}{(t_2+t_4)N_0}\right).
\end{equation}
Similarly, the achievable data rate of user $Y$ to $X$ in $t_3$ is
{\small
\begin{equation}
\label{Ry3}
R_Y^{(3)}=t_3\log_{2}\left(1+\frac{E_Yh_{YX}}{(t_3+t_4)N_0}\right).
\end{equation}
}
In the $4$-th time slot, as the transmitter channel state information (CSI) is not available at the two users, we use Alamouti space-time block code (STBC) transmit diversity scheme \cite{Alamouti} for joint information transmission with $t_4^{(1)} = t_4^{(2)}$. With transmit power of $P_X$ and $P_Y$ for the two users, the achievable data rate of user $X$ in the time slot is
\begin{equation}
\label{Rx4}
\small
R_X^{(4)}=t_4/2\log_{2}\left(1+ \frac{E_Xh_{XD}}{(t_2+t_4)N_0} + \frac{E_Yh_{YD}}{(t_3+t_4)N_0} \right ),
\end{equation}
and $R_Y^{(4)}=R_X^{(4)}$ for user $Y$.

Notice that the DN can overhear the transmission of user $X$ ($Y$) in the $2$-nd ($3$-rd) time slot, although not dedicated to it. In theory, the DN can improve the data rates of both users with the overheard signals \cite{2014:Ju3}. In this paper, however, we consider a practical coding scheme and a simple receiver structure of DN so that the DN only decodes each user's information transmission in the $4$-th time slot instead of performing joint decoding from the signal received in two time slots. The case with joint decoding will be considered as a future work. In this case, the achievable rates of user $X$ and $Y$ are
\begin{equation}\label{}
\small
  R_X = \min \left(R_X^{(2)},R_X^{(4)}\right),\ R_Y = \min \left(R_Y^{(3)},R_Y^{(4)}\right),
\end{equation}
where $R_X^{(2)}$ and $R_Y^{(3)}$ are in (\ref{Rx2}) and (\ref{Ry3}), respectively, while $R_X^{(4)}$ and $R_Y^{(4)}$ are in (\ref{Rx4}).

In WPCNs, the user data rates can differ significantly, e.g., by two orders of amplitude, because of the disparities in both energy harvesting performance and information transmit power consumptions. As a common indicator of throughput fairness (e.g., see \cite{2014:Ju1}), we adopt the minimum throughput of the two users as the performance metric, i.e.,
\begin{equation}
\label{Rmaxmin}
\small
R=\min(R_X,R_Y).
\end{equation}
In particular, we are interested in the following optimal time allocation problem to maximize the throughput performance:
{\small
\begin{equation}
   \begin{aligned}
    &\max_{t_1,t_2,t_3,t_4} & &  \min(R_X,R_Y)\\
    &\text{s. t.}    & & t_1+t_2+t_3+t_4 = 1 - t_0, \label{P1}\\
    & & & t_1,t_2,t_3,t_4\geq 0.
   \end{aligned}
\end{equation}}

\section{Optimal Throughput of the Proposed User Cooperation Method}
In this section, we study the optimal throughput performance of the considered user cooperation scheme by proposing efficient algorithm to solve (\ref{P1}).

\subsection{Analysis of Optimal Solution}
To begin with, we first show that the optimal solution to (\ref{P1}) should allow the two terminal users to transmit at an equal rate, i.e., $R^*_X =R^*_Y$. Otherwise, if $R_X^* \neq R_Y^*$, we assume without loss of generality that $R_X >R_Y$, and the case with $R_X<R_Y$ follows. In this case, $R_X^{(2)}>R_Y^{(3)}$ must hold because $R_X^{(4)} = R_Y^{(4)}$. Note that given a pair of $(t_1,t_4)$, $R_X^{(2)}$ ($R_Y^{(3)}$) is an increasing (a decreasing) function of $t_2$, for $t_2 \in\left[1-t_0-t_1-t_4\right]$, which is proved in Lemma 3.1 and demonstrated numerically in Fig. 2(a). Accordingly, we can always adjust $t_2$, and thus $t_3$, to improve the objective of (\ref{P1}) until $R_X^{(2)}=R_Y^{(3)}$. Therefore, we can conclude that $R_X^* =R_Y^*$ must hold for the optimal solution of (\ref{P1}). Accordingly, the optimal throughput in (\ref{P1}) is often referred to as \emph{common throughput} \cite{2014:Ju1}.

\underline{\emph{Lemma}} \emph{3.1}: $R_X^{(2)}$ increases monotonically and $R_Y^{(3)}$ decreases monotonically in $t_2 \in\left[0,T_0\right]$, where $T_0= t_2 +t_3$ is a fixed parameter.

\emph{Proof:}
Please refer to Appendix A.

Then, we show that $R_X^{(2)*}=R_X^{(4)*}$. Otherwise, if $R_X^{(2)*}>R_X^{(4)*}$ (or $R_X^{(2)*}<R_X^{(4)*}$), we can easily increase the objective in (\ref{P1}) by allocating more time on WET, and less time for user cooperation in $t_2$ and $t_3$ (or joint transmission in $t_4$). Similarly, we have $R_Y^{(3)*}=R_Y^{(4)*}$. Based on the above analysis, we conclude that the optimal solution must satisfy
\begin{equation}\label{P2}
\small
  R_X^{(2)*}=R_Y^{(3)*}=R_X^{(4)*},
\end{equation}
where $R_X^{(4)*}=R_Y^{(4)*}$ holds because of the Alamouti STBC in use. We can express the terms in (\ref{P2}) as functions of time allocation as following
{\small
\begin{align}
R_X^{(2)}(t_1,t_2,t_4)&=t_2\log_{2}\left(1+ \rho_1 \frac{t_1}{t_2+t_4}\right) ,  \label{rx2} \\
R_Y^{(3)}(t_1,t_3,t_4)&=t_3\log_{2}\left(1+ \rho_2 \frac{t_1}{t_3+t_4}\right) ,   \label{ry3}\\
R_X^{(4)}(t_1,t_2,t_3,t_4)&=t_4/2\log_{2}\left(1+ \rho_3 \frac{t_1}{t_2+t_4}+ \rho_4 \frac{t_1}{t_3+t_4}\right) ,\label{rx4}
\end{align}
}
where ${\rho_1}\triangleq{ \eta P_t h_{EX} h_{XY} } / {N_0}$, ${\rho_2}\triangleq{ \eta P_t h_{EY} h_{YX} } / {N_0}$, ${\rho_3}\triangleq{ \eta P_t h_{EX} h_{XD}}/ {N_0}$, and ${\rho_4}\triangleq{ \eta P_t h_{EY} h_{YD}}/ {N_0}$ are parameters.

\begin{figure}
  \centering
  \begin{center}
    \includegraphics[width=0.4\textwidth]{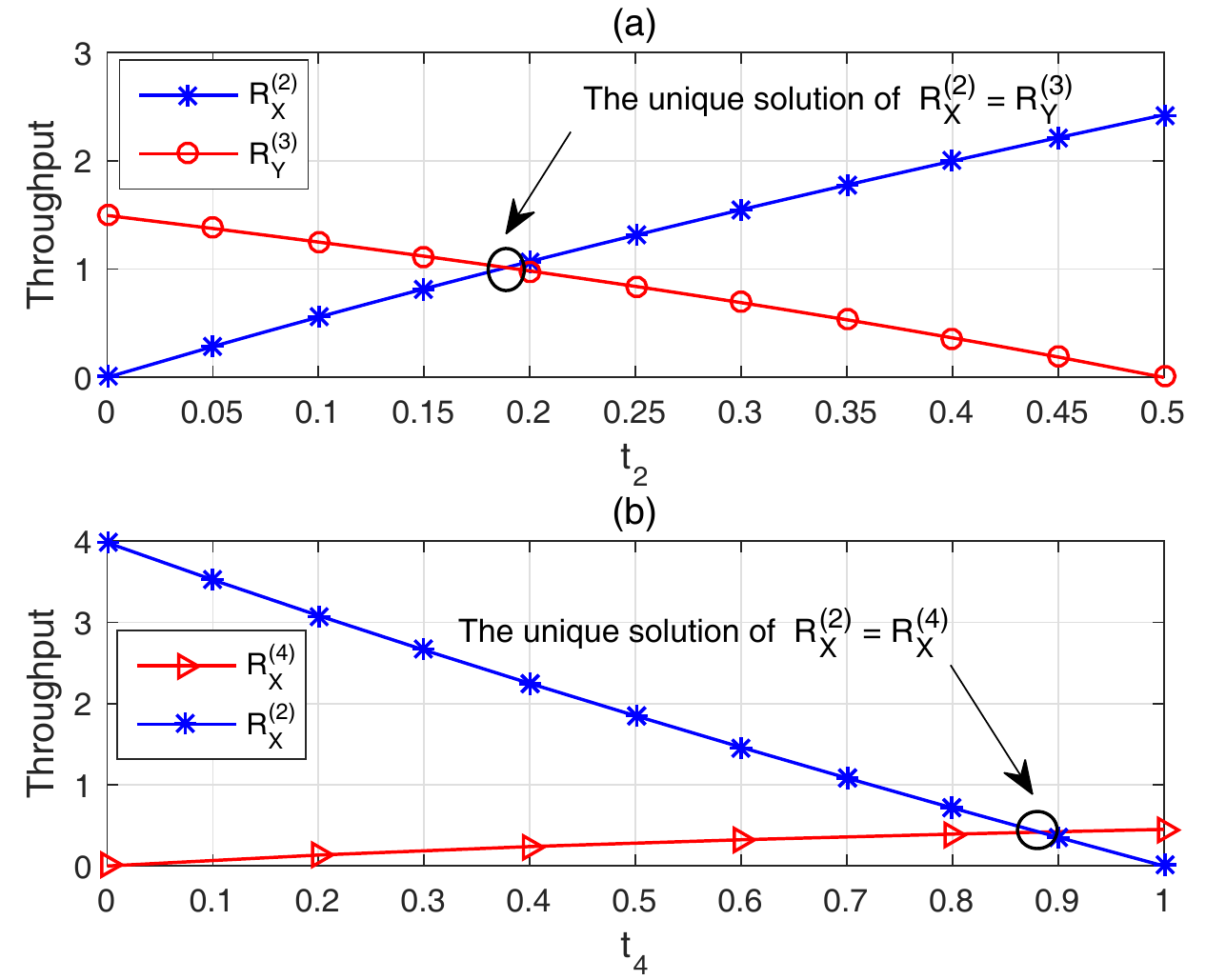}
  \end{center}
  \caption{Numerical results of the monotonic properties in Lemma 3.1 (sub-figure (a) above) and Lemma 3.2 (sub-figure (b) below). The detailed system parameters are specified in Section IV. }
  \label{Fig.2}
\end{figure}

\begin{algorithm}
\footnotesize
 \SetAlgoLined
 \SetKwData{Left}{left}\SetKwData{This}{this}\SetKwData{Up}{up}
 \SetKwRepeat{doWhile}{do}{while}
 \SetKwFunction{Union}{Union}\SetKwFunction{FindCompress}{FindCompress}
 \SetKwInOut{Input}{input}\SetKwInOut{Output}{output}
 \Input{time duration $T=1$, channel estimation time $t_0$}
 \Output{the optimal time allocation of $\left\{t_1^*,t_2^*,t_3^*,t_4^*\right\}$}
 Initialize: $t_1\leftarrow 0$, $R^*\leftarrow 0$, $\Delta \leftarrow$ small positive step size\;
   \While{$t_1 \leq 1-t_0$}{
 $t_1 \leftarrow t_1+\Delta$\;
    $UB_4 \leftarrow  1-t_0-t_1$, $LB_4 \leftarrow 0$\;
    \Repeat{$|R_X^{(2)}-R_X^{(4)}|<\sigma$}{
    $t_4\leftarrow \left(UB_4+LB_4\right)/2$\;
    $UB_2 \leftarrow  1 - t_0 - t_1 -t_4$, $LB_2 \leftarrow 0$\;
    \Repeat{$|R_X^{(2)}-R_Y^{(3)}|<\sigma$}{
    $t_2\leftarrow \left(UB_2+LB_2\right)/2$,\;
    $t_3\leftarrow 1-t_0-t_1-t_4-t_2$\;
    Calculate $R_X^{(2)}$ and $R_Y^{(3)}$ using (\ref{rx2}) and (\ref{ry3}), respectively\;
    \eIf{$R_X^{(2)}>R_Y^{(3)}$}{
    $UB_2 \leftarrow t_2$;}
    {
    $LB_2 \leftarrow t_2$;
    }
    }

    Given $t_2,t_3,t_4$, calculate $R_X^{(4)}$ using (\ref{rx4})\;
    \eIf{$R_X^{(4)}>R_X^{(2)}$}{
    $UB_4 \leftarrow t_4$;}
    {
    $LB_4 \leftarrow t_4$;
    }
    }
    $R\leftarrow \min\left(R_X^{(2)},R_X^{(4)}\right)$\;
    \If{$R>R^*$}{
    $R^*\leftarrow R$, \ \
    $\left\{t_1^*,t_2^*,t_3^*,t_4^*\right\}\leftarrow \left\{t_1,t_2,t_3,t_4 \right\}$\;
    }

}
\textbf{Return} $\left\{t_1^*,t_2^*,t_3^*,t_4^*\right\}$.
\caption{Optimal time allocation solution to (\ref{P1}).}
\label{alg1}
\end{algorithm}

\subsection{Optimal Solution Algorithm to (\ref{P1})}
Based on the above analysis, we proposed in this subsection an efficient algorithm to solve (\ref{P1}). To begin with, we show that there always exists a unique time allocation $(t_2,t_3,t_4)$ that satisfies (\ref{P2}) given a fixed $t_1$. To see this, from Lemma 3.1, we can always find a unique set of $(t_2,t_3)$ to satisfy $R_X^{(2)} = R_{Y}^{(3)}$ when a set of $(t_1,t_4)$ is given, such that $t_2+t_3 = 1-t_0-t_1-t_4$ is a fixed parameter. Equivalently, we can denote $R_X^{(2)}$ and $R_Y^{(3)}$ as functions of $(t_1,t_4)$, i.e., $R_X^{(2)}(t_1,t_4)$ and $R_Y^{(3)}(t_1,t_4)$, respectively. Besides, $R_X^{(4)}$ can also be expressed as a function $(t_1,t_4)$, i.e., $R_X^{(4)}(t_1,t_4)$, because $(t_2,t_3)$ is uniquely determined by a pair of $(t_1,t_4)$. Then, given a fixed $t_1$, a unique $t_4$ can be found to satisfy $R_X^{(2)}(t_1,t_4) = R_X^{(4)}(t_1,t_4)$, because $R_X^{(2)} = 0$ when $t_4 = 1-t_0-t_1$ and $R_X^{(2)}$ decreases with $t_4 \in\left[0,1-t_0-t_1\right]$, while $R_X^{(4)} = 0$ when $t_4 = 0$ and $R_X^{(4)}$ increases with $t_4 \in\left[0,1-t_0-t_1\right]$. In particular, given a fixed $t_1$, the monotonic properties of $R_X^{(2)}(t_1,t_4)$ and $R_X^{(4)}(t_1,t_4)$ as a function of $t_4$ is proved in the following Lemma 3.2 and illustrated numerically in Fig.~\ref{Fig.2}(b).

\underline{\emph{Lemma}} \emph{3.2}: $R_X^{(4)}$ increases monotonically and $R_X^{(2)}$ decreases monotonically in $t_4\in \left[0,T_1\right]$, where $T_1=t_2+t_3+t_4$ is a fixed parameter.

\emph{Proof:}
Please refer to Appendix B.

Now that a unique time allocation $(t_2,t_3,t_4)$ can be found with a fixed $t_1$, the optimal solution to (\ref{P1}) can be obtained by a simple line search over $t_1\in \left[0,1-t_0\right]$. The key idea is that, given a pair of $(t_1,t_4)$, the unique set of $(t_2,t_3)$ that satisfies $R_X^{(2)} = R_{Y}^{(3)}$ can be obtained by a bi-section search over $t_2\in\left[0,1-t_0-t_1-t_4\right]$. Accordingly, we can calculate $R_X^{(2)}$ and $R_X^{(4)}$ using (\ref{rx2}) and (\ref{rx4}), respectively, based on which we can find a unique $t_4$ that satisfy $R_X^{(2)}(t_1,t_4) = R_X^{(4)}(t_1,t_4)$ using a bi-section search over $t_4\in[0,1-t_0-t_1]$. Then, we only need to perform a linear search over $t_1\in[0,1-t_0]$ to find the optimal set of $(t_1,t_2,t_3,t_4)$ that produces the largest common throughput. A pseudo-code of the above searching algorithm is summarized in Algorithm 1, where the lines $7-17$ correspond to the bi-section search over $t_2$ and lines $4-24$ correspond to the bi-section search over $t_4$. The time complexity of the algorithm is proportional to $1/\Delta \cdot \left[\log(1/\sigma)\right]^2$, where $\Delta$ and $\sigma$ are small positive parameters determined by the solution precision requirement. The proposed algorithm is of low complexity, which enables fast calculation of the optimal time allocation solution.

\subsection{Benchmark Methods}
For performance comparison, we consider in Fig.~3 two benchmark methods: the two users do not cooperate and transmit to the DN in a TDMA manner (Non-cooperate); and one user acts as the relay for the other (Relay) \cite{2014:Ju3,2015:HeCHen,2015:Nasir}. For fair comparison, we use the same channel estimation method for both the Non-cooperate and Relay schemes as the proposed user cooperation method. Accordingly, CE consumes the same amount of time $t_0$. For both methods, the first time slot $t_1$ is assigned for WET and the remaining time is used for WIT. For the Relay scheme, the WIT time is divided into two time slots $t_2$ and $t_3$. During $t_2$, one user uses the harvested energy to transmit target information to another. In $t_3$, another user will help forward the information received in $t_2$ and transmit its own information to the DN. In particular, either user can act as the relay for the other (i.e., $Y{\rightarrow}X{\rightarrow}D$ and $X{\rightarrow}Y{\rightarrow}D$). In this paper, we choose the better one between the two scenarios in different conditions to represent the Relay scheme. Different from Cooperate and Relay scheme, user $X$ and $Y$ of the Non-cooperate scheme transmit their independent information to the DN directly in $t_2$ and $t_3$, respectively. For fair comparisons, we assume that all the nodes exhaust their harvested energy and transmit with constant power in the WIT stage, and the DN cannot perform joint decoding for a user's message received from different time slots that are in different data rates. Here, the methods to optimize the time slot allocation for the benchmark schemes are omitted due to the page limit.

\begin{figure}
  \centering
   \begin{center}
      \includegraphics[width=0.4\textwidth]{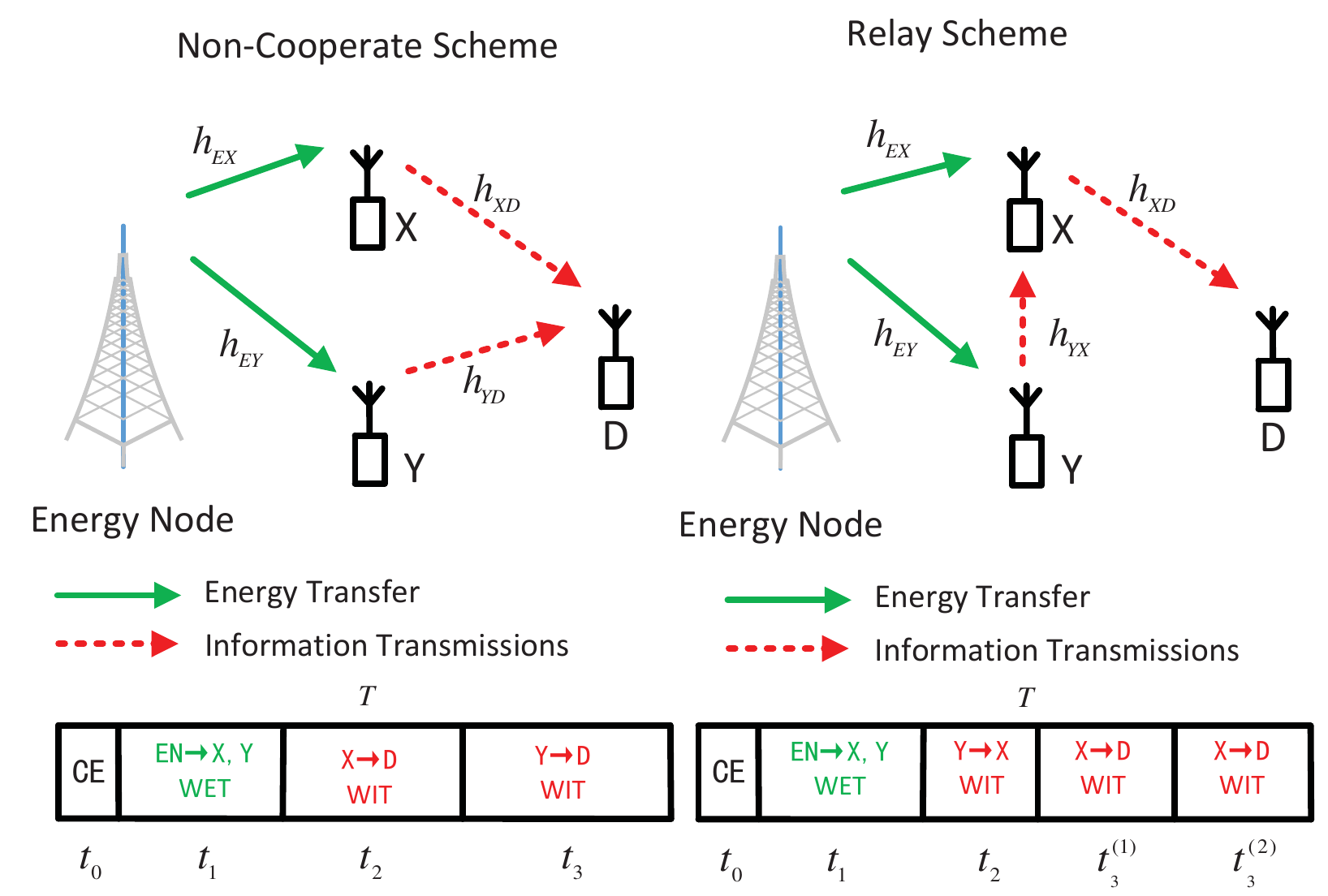}
   \end{center}
  \caption{Illustration of the two benchmark methods considered.}
  \label{Fig.3}
\end{figure}

\section{Simulation Results}
In this section, we evaluate the performance of the proposed user cooperation under different channel conditions. In all figures, the optimal common throughput performance of different schemes are presented. Unless otherwise stated, it is assumed that the distance between the EN and user $X$ and $Y$ is $5$m and $10$m, respectively, and the two users are separated by $2$m. The transmit power of EN is $P_t=3$W, the wireless channel gain $h_{ij}=(\frac{3\cdot10^8}{4{\pi}df_d})^{d_{D}}$, $ij\in\{EX,EY,XY,YX,XD,YD\}$, where $f_d$ denotes $915$MHz carrier frequency, and $d_{D} = 2$ denotes the path loss exponent. The antenna power gain is $2$.

Fig.~4 shows the impact of user-to-DN channel disparity to the optimal common throughput performance. Without loss of generality, we fix $h_{YD}=4.25\times10^{-7}$ (this corresponds to the average channel gain when $Y$ is $40$ meters from the DN) as a constant and show the performance when $h_{XD}$ becomes smaller. Note that when $h_{YD}/h_{XD}$ changes from $0$ to $5 $dB, the common throughput of the Non-cooperate scheme hardly changes while those of the Cooperate scheme and Relay scheme decrease more evidently. For the Non-cooperate scheme, this is because the $0-5$ dB case corresponds to the energy-constrained region, where the major performance bottleneck is the less energy harvested by $Y$ due to the poor EN-to-Y channel. Therefore, moderate decrease of user $X$'s data rate will not change the common throughput performance. For the Relay and user cooperation schemes, however, the data rate performance is more sensitive to the channel degradation of X-to-DN channel, as it needs to transmit the messages of both two users. Besides, the Relay scheme switches from user $X$ being the relay to user $Y$ being the relay when $h_{YD}/h_{XD} > 6$ dB. We can see that the proposed Cooperate scheme outperforms the other two schemes in almost all the scenarios, expect for a minor performance loss compared to the Non-cooperate scheme when $h_{YD}/h_{XD} = 6$ dB. Nonetheless, the performance of the Non-cooperate scheme degrades drastically as $h_{XD}$ further decreases. The proposed user cooperation and the Relay method perform comparable when $h_{XD}$ becomes very small, as most data is now sent from user Y to the DN. Fig.~4 shows that proposed cooperation method is robust against user-to-DN channel disparity because of the channel diversity achieved in transmitting user messages.

\begin{figure}
  \centering
   \begin{center}
      \includegraphics[width=0.4\textwidth]{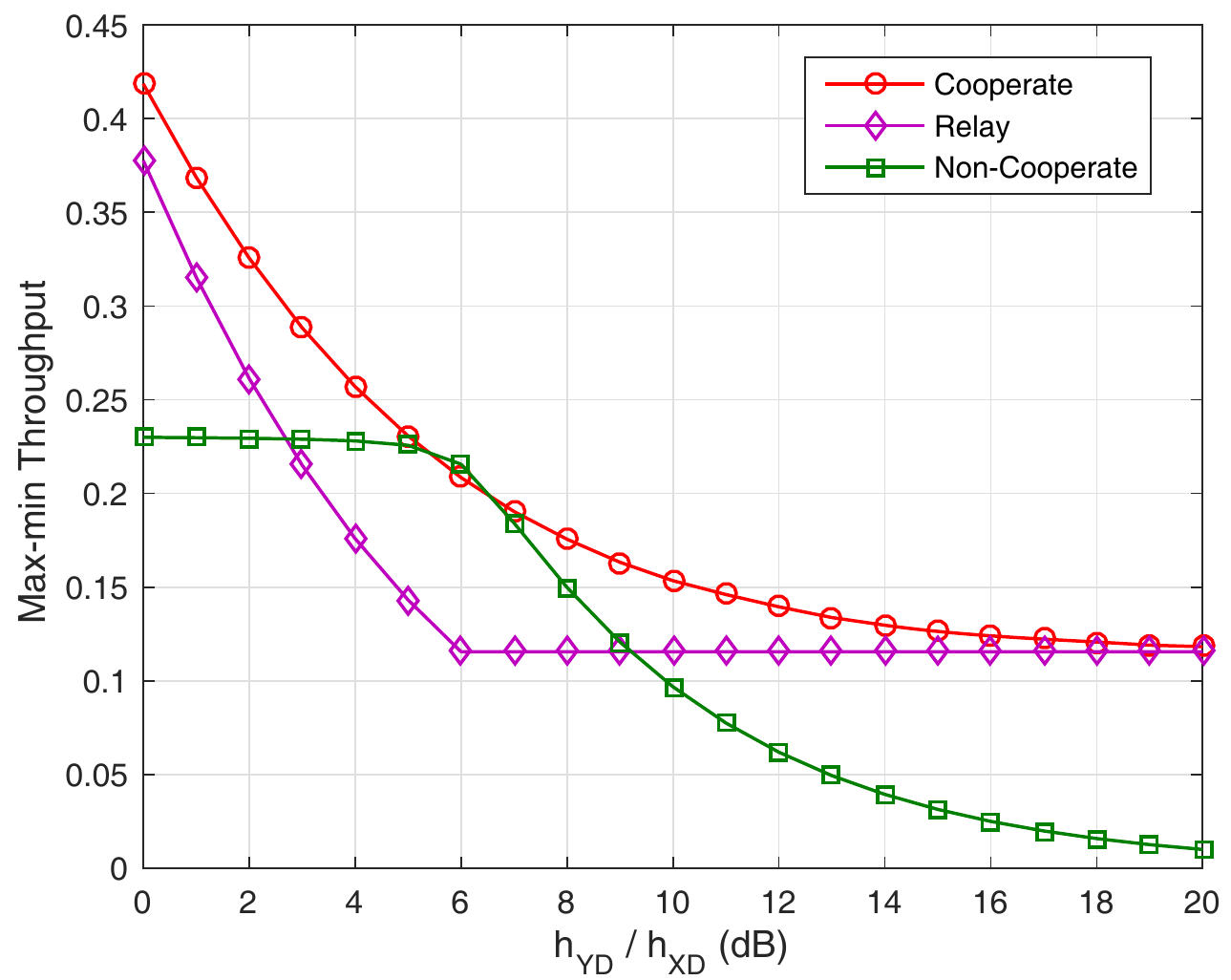}
   \end{center}
  \caption{The impact of user-to-DN channel disparity to the optimal common throughput performance.}
  \label{Fig.4}
\end{figure}

Fig.~5 shows the impact of EN-to-user channel disparity to the optimal common throughput performance. Here, we set $h_{XD}=h_{YD}=4.25\times10^{-7}$, fix $h_{EX}=2.72\times10^{-5}$ as a constant and show the performance when $h_{EY}$ becomes smaller. Notice that when $h_{EX}/h_{EY}=1$, i.e., the two users harvested the same amount of energy, the Non-cooperate scheme is slightly better than the Cooperate scheme due to the extra information exchange time consumed by the proposed cooperation scheme. However, as the differences between $h_{EX}$ and $h_{EY}$ increases, the Cooperation scheme degrades moderately while the Non-cooperate scheme degrades significantly. It is worth noting that the performance of the Relay scheme ($Y{\rightarrow}X{\rightarrow}D$) changes marginally compared other two schemes as $h_{EX}$ changes. This is because the distance between two users is much shorter than the EN-to-user distance. In this case, the change in the energy harvested by user $Y$ has marginal impact on the throughput. The results in Fig.~5 demonstrate the superior performance of the proposed user cooperation method, thanks to the energy diversity achieved from allowing the users to share their energy to transmit jointly their messages.
\begin{figure}
  \centering
   \begin{center}
      \includegraphics[width=0.4\textwidth]{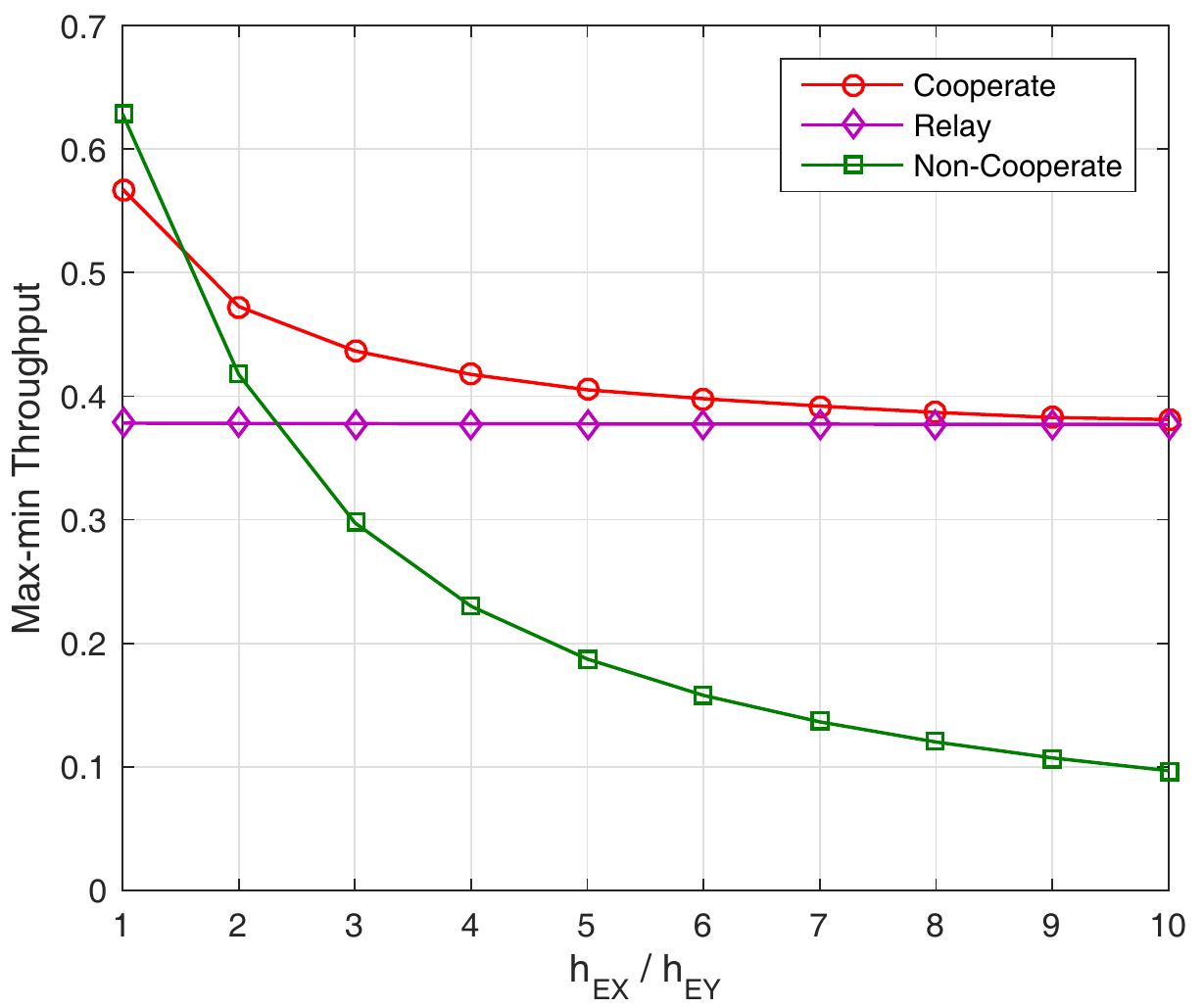}
   \end{center}
  \caption{The impact of EN-to-user channel disparity to the optimal common throughput performance.}
  \label{Fig.5}
\end{figure}
\begin{figure}
  \centering
   \begin{center}
      \includegraphics[width=0.4\textwidth]{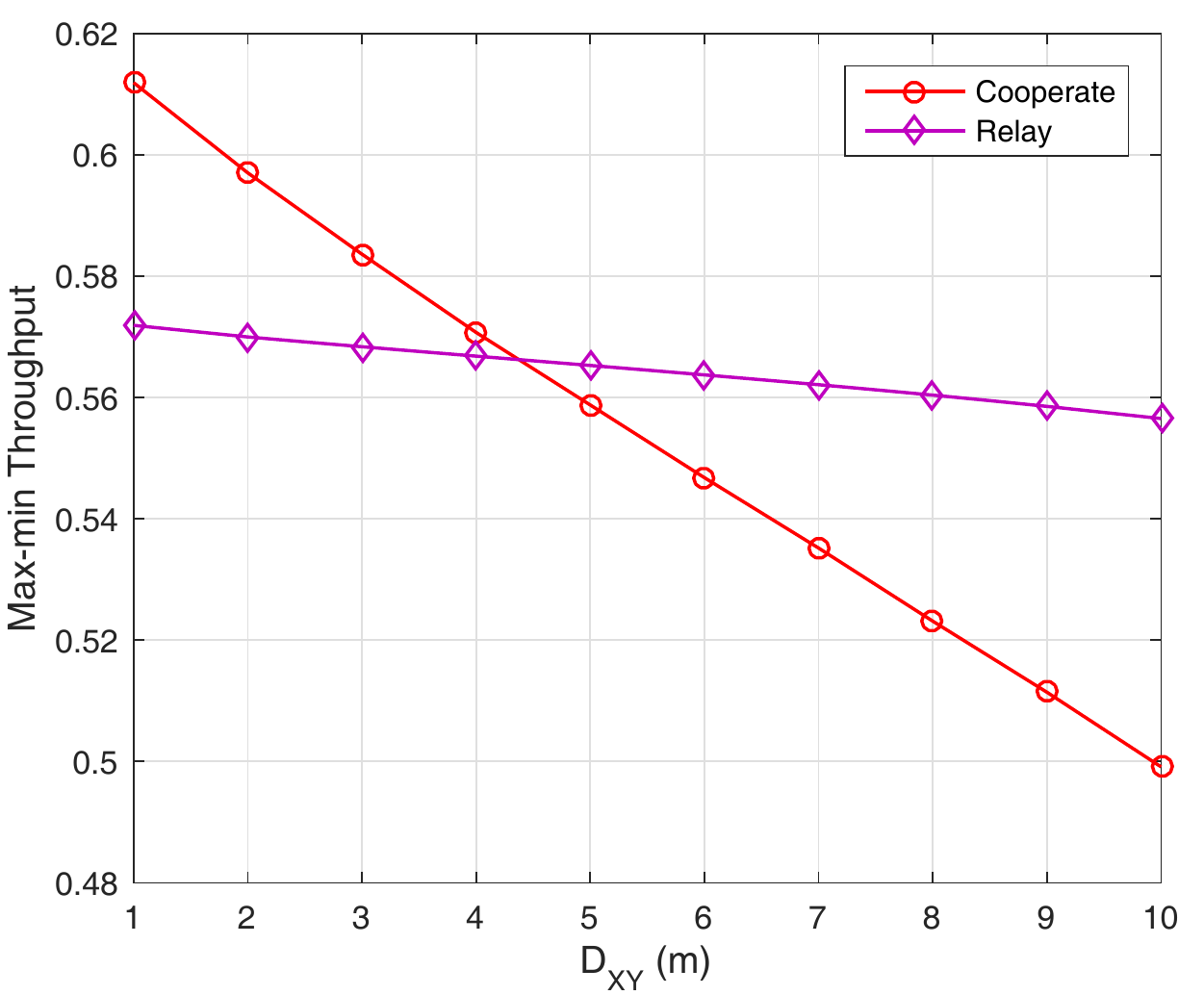}
   \end{center}
  \caption{The impact of inter-user channel strength to the optimal common throughput performance.}
  \label{Fig.6}
\end{figure}

In addition, Fig.~6 shows the impact of inter-user channel strength to the optimal common throughput performance. Here, we set $h_{XD}=h_{YD}$ and $h_{EX}=4h_{EY}=2.72\times10^{-5}$, and vary the distance between user $X$ and $Y$ from $1$m to $10$m. As the performance of Non-cooperate scheme is independent of $D_{XY}$, we only compare the Cooperation scheme and Relay scheme. It is observed that the max-min throughput of both schemes decreases with $D_{XY}$. However, the Cooperation scheme is more sensitive to the channel degradation between the cooperating users than the Relay scheme because it has to use this channel twice during information exchange while the Relay scheme only needs once. We can therefore conclude that user cooperation is most effective when the inter-user channel is sufficiently strong to support efficient user message exchange.

It is also worth noting that no scheme is optimal in all scenarios. In general, different scheme should be applied based on the network setups and parameters. However, the proposed user cooperation method has demonstrated superior performance under different setups, especially when two users are close with each other and have small discrepancies on channel condition to the DN. Thanks to the channel and energy diversity gains achieved, the proposed cooperation method shows robust performance under most scenarios.

\section{Conclusion}
This paper studied a two-user WPCN in which a new user cooperation method is exploited to improve the throughput fairness. We derived the maximum common throughput achieved by the proposed user cooperation and performed numerical analysis to study the impact of system setups to the throughput performance. By comparing with two representative benchmark methods, we showed that the proposed user cooperation method can effectively achieve both channel and energy diversity gains to enhance the throughput fairness under different setups.

\appendices
\section{Proof of Lemma 3.1}
The transmit power of user $X$ is $P_X=E_X/(t_2+t_4)$, we have from (\ref{Rx2}) that
\begin{equation}\label{proof3.1}
\small
R_X^{(2)}= t_2\log_{2}\left(1+\frac{E_Xh_{XY}}{(t_2+t_4)N_0}\right) \triangleq t_2\log_{2}\left(1+ \frac{c_1}{t_2+c_2}\right),
\end{equation}
where $c_1\triangleq E_Xh_{XY}/N_0$, $c_2 \triangleq t_4$ are both constant. By taking the first and second order derivatives of $R_X^{(2)}$ in $t_2$, we have
{\small
\begin{align}\label{}
\frac{dR_X^{(2)}}{dt_2} & =\log_{2}\left(1+ \frac{c_1}{t_2+c_2}\right)- \frac{c_1t_2}{\ln2(t_2+c_3)(t_2+c_2)}, \\
\frac{d^2R_X^{(2)}}{dt_2^2} & = - \frac{c_1}{\ln2}  \frac{(c_2+c_3)t_2+2c_2c_3}  {(t_2+c_3)(t_2+c_2)},
\end{align}
}
where $c_3\triangleq c_1+c_2$. Because $\frac{d^2R_X^{(2)}}{dt_2^2}<0$ and $ \lim\limits_{t_2 \to +\infty} \frac{dR_X^{(2)}}{dt_2} = 0 $, we can infer that $ \frac{dR_X^{(2)}}{dt_2} > 0 $ when $t_2>0$, which leads to the proof of Lemma 3.1 that $R_X^{(2)}$ increases in $t_2 \in\left[0,T_0\right]$. Similarly, we have $R_Y^{(3)}$ deceases with $t_2 \in \left[0,T_0\right]$.  $\hfill \blacksquare$

\section{Proof of Lemma 3.2}
First of all, we show that both $t_2$ and $t_3$ decrease as $t_4$ increases. Otherwise, we assume without loss of generality that $t_2$ increases and $t_3$ decreases when $t_4$ become larger. We denote the updated values of $t_2$ and $t_3$ after $t_4$ becomes $\bar{t}_4=t_4+{\Delta}t_4$ as $\bar{t}_2=t_2+{\Delta} t_2$ and $\bar{t}_3=t_3-{\Delta} t_3$, respectively, where ${\Delta} t_2,{\Delta} t_3,{\Delta}t_4>0$, and ${\Delta} t_2+ {\Delta}t_4 -{\Delta} t_3 =0$. Besides, we denote the updated values of $R_X^{(2)}$ and $R_Y^{(3)}$ as $\bar{R}_X^{{(2)}}$ and $\bar{R}_Y^{{(3)}}$, respectively. It can be easily shown from Lemma 3.1 that $\bar{R}_X^{{(2)}}>\bar{R}_Y^{{(3)}}$ given $R_X^{(2)} = R_Y^{(3)}$. However, this contradicts with the necessary condition of an optimal solution that requires $\bar{R}_X^{{(2)}}=\bar{R}_Y^{{(3)}}$. Therefore, we reject our assumption and conclude that both $t_2$ and $t_3$ decrease as $t_4$ increases. Because $t_2+ t_3 +t_4 =T_1$, we can infer that $t_2+t_4 = T_1 -t_3$ increases with $t_4$, so does $t_3+t_4$. This, together with the result that $t_2$ (and $t_3$) decrease with $t_4$, leads to the proof that $R_X^{(2)}$ in (\ref{rx2}) (and $R_Y^{(3)}$ in (\ref{ry3})) is a decreasing function with $t_4$. Besides, we can also show that $R_X^{(4)}$ in (\ref{rx4}) increases with $t_4$ by calculating the partial derivatives of $R_X^{(4)}$ over the vector $(t_2,t_3,t_4)'$, which is omitted here due to the page limit. $\hfill \blacksquare$

%\ifCLASSOPTIONcaptionsoff
%  \newpage
%\fi

%\end{IEEEbiography}

\end{document}